\documentclass[12pt,a4paper]{article}
\usepackage{graphicx}
\renewcommand{\baselinestretch}{1.4}
\date{}
\begin{document}
\title{ \sc Dynamical properties of nystagmus}
\author{Nikolay K. Vitanov  \thanks{corresponding author} and Veneta Todorova}
\maketitle
\begin{abstract}
\small
By the methods of the time series analysis we investigate human vestibular
and optovestibular nystagmus time series. They can be stationary or nonstationary  and
are between the classes of  purely periodic or purely chaotic time series.
The amplitudes of the autocorrelation functions of the time series have
an unexpected peak between $7$ and $9$ seconds on the
time axis. The singular spectrum analysis
shows that the important information about the time series is concentrated
in the first four to six principal components. The influence of the
sensory input modality, which can be changed by opening or closing eyes of the
investigated persons, leads to changes in the histogram, but leaves almost
unchanged other important characteristics of the time series such as the
statistical dimension or the positions of the dominant frequencies in the
power spectrum. Thus the behavior of the studied time series shows a presence of
nonlinearities which could play an important role for realization of the
nystagmus phenomenon.
\end{abstract}
\begin{flushleft}
\end{flushleft}
\small
\par
Nystagmus [$^{1}$] is a  rhythmic involuntary eyes movement that stabilizes
gaze in the space during head movements or rotations of the environment.
By means of this physiological mechanism the images on the
retina do not move with respect to the main optic coordinates and a clear
vision is maintained. The nystagmus reactions are part of the general
function with critical biological importance - orientation in the space.
These reactions are very important for health organisms under the
action of extreme influences such as the sport, astronautics,
sailing, construction of high buildings, high-speed transport, etc. In all
of these cases the professional selection based on medical examination
include as a main factor the nystagmus reactivity.
\par
Many structures are involved in the realization of
the nystagmus reaction: from vestibular and optovestibular receptor structures
to nervous-oculomotor apparatus with involving of large brain zones
(from brain-stem to cortex) in the sensory-motor integration.
As a final product the nystagmus
is a carrier of important information for the functional status of the
forming structures. This makes the nystagmus an obligatory criterion in the
clinical diagnostic of neurological, vestibular and eyes disfunctions.
The  nystagmus eyes movements are to the same amount
but in the opposite direction with respect to the direction of the
movement of the head. The nystagmus  has two alternative
phases of slow and fast movements in opposite directions. The slow movement
phase  is connected to the graceful following eye movements and contains
the biological part of the nystagmus reflex. The fast movement phase
is connected to the jump-like saccades and reflects only the mechanical
reposition of the eyes. The vestibular and the
visual system ensure the afference of the nystagmus called vestibular or
optovestibular in dependence on the sensory input modality. With respect to its
external manifestation the two kinds of the nystagmus do not differ each from
another.
\par
The goal of this paper is to investigate characteristic peraccelatory
nystagmus time series by the methods of the
nonlinear time series analysis [$^{2}$,$^{3}$,$^{4}$].  Despite the progress in the area of the
registration and evaluation techniques all characteristic nystagmus
quantities have large deviations from its assumed physiological normal
values. Thus it is very difficult to distinguish between the norm and the
pathology. This makes the problem for the investigating of the nystagmus
highly actual today. The experimental data are chosen among
the time series, obtained under physiological conditions from a
healthy volunteer, a man aged 32 years, without history of past or present vestibular,
otologic, opthalmologic or neurologic diseases, that potentially could affect
the nystagmus results. For the experiment we have used an electronically
controlled rotating chair (Toennies, Freiburg). Stimuli consisted of
acceleration steps ($\pm 8^{\circ} /s^{2}$, $190^{\circ} /s$) to the right 
and to the left.
Rotation test have been done under two different visual conditions:
in complete darkness with eyes closed (vestibular nystagmus), and with
eyes open in a lighted optokinetic drum (optovestibular
nystagmus). During passive body rotation in the light concomitant visual and
vestibular information interact in order to provide better ocular stability
[$^{5}$]. Horizontal eye movements have been recorded bipolarly from surface
electrodes at the outer canti by means of conventional electronystagmography.
It is based on the registration of the biopotential
difference which arises between the retina and the cornea in the cases of different
movement of the eyes.  The eye-ball is a dipole: the one pole, the cornea is
positive charged, and  the another pole, the retina is negative charged.
When the human sees in the straight direction the potential between the
cornea and the retina is small. When the sight changes its direction this
potential can increase ten times. From the registered nystagmus reaction we can estimate the
basic nystagmus parameters as length in the time, frequency. amplitude and
velocity for the different components of the nystagmus cycle as well as for the
entire nystagmus reaction. Normally one calculates the velocity of the slow
phase of the nystagmus. It is a functional analogy of the compensatory
eye-ball movements.
\par
The nystagmus time series can be stationary
or nonstationary ones. This is shown in panels {\it a} and {\it b} of Fig. 1.
The time series have complicated structure because of the influence of afference
sense structures, effectors occumotor apparatus and connecting segments
of the central nervous system. The stationary time series for vestibular
nystagmus are shown in panel {\it a}. The time series connected to the optovestibular
nystagmus  shown in panel {\it b} are clearly nonstationary with a trend given by
the relationship $V=8.3 -0.045 t$, where $V$ is the angular deviation of
the eye-balls from the strait sight direction in the horizontal plane and $t$ is the time.
After a detrending these time series become stationary one and we shall use this stationary form
 in the analysis below. We perform  time delay embedding of the time
series, obtaining the value of the delay ($\tau=4$) and the minimum embedding
dimension  ($m=4$) by means of the first minimum of the mutual information
[$^{6}$] and the method of the false nearest neighbors [$^{7}$].
The panels {\it c} and {\it d} of  Fig. 1 show the results.
The obtained manifolds contain  points  scattered almost on one plane and we
can expect that the correlation dimensions $D_{2}$ [$^{8}$, $^{9}$] of 
the investigated
time series are between $1$ and $2$. Indeed $D_{2}=1.3 \pm 0.14$ for the
vestibular nystagmus time series and $D_{2}=1.4 \pm 0.12$ for the optovestibular
nystagmus time series. Thus the change of the influence of the environment
do not change quite significantly the value of the correlation dimension of the
time series and they remain in the middle class between purely periodic and
purely chaotic motions in the phase space.
\par
Fig. 2 shows the autocorrelations, histograms and the lower parts of the
power spectra for the investigated time series. The autocorrelation
function exhibits a periodic behavior with  large region of almost exponential decaying of the
extrema of $A(\tau)$. After this the correlation of the data increases,
and the beginning of the regions of increasing are almost on the
same place in the two time series. The autocorrelations of periodic
processes are periodic and the autocorrelations for chaotic processes
decay exponentially . The autocorrelations of the investigated time series
show exponential decaying of the amplitude of the maxima for small $\tau$
and thus are close to the chaotic systems. But between $350$ and $450$
sampling units we observe increasing of amplitude of
the autocorrelations for both time series. This peak  between $7$
and $9$ seconds of time hints that the nystagmus
dynamics is nonlinear one. For stationary
time series the power spectrum does not change in course of evolution
of the time series.  The power spectrum does not show the famous $1/f$-
behavior. The histograms shown in panels {\it c} and{\it d} of Fig. 2
show that the distribution of the values of the
time series is far from the Gaussian one. The vestibular nystagmus data exhibit
fat tails and the optovestibular data histogram has two peaks. The
power spectral density is concentrated in the same two frequencies for the
vestibular and for the optovestibular nystagmus. Thus the changing environmental
conditions do not change the dominant frequencies of the nystagmus dynamics.
\par
The singular spectrum  and the principal components of the time series
are convenient tool for analysis of stationary and nonstationary time series
[$^{10}$, $^{11}$]. The largest singular values of the investigated
time series are connected to their periodic components. The
contribution of each principal component to the total variance of the time
series is a measure of the important information about the series which is
stored in the correspondent principal component. Thus by means of the
variance we can determine the number of the important principal components
and we are able to construct  on their basis noise-free time series, which have all
important features of the original series. The number of the important
singular values gives the statistical dimension of the time series which is
an upper bound on the correlation dimension. The results of the singular
spectrum analysis performed in twenty-dimensional phase space of vectors,
obtained from the time series by means of unit delay, are presented in Fig. 3.
Panels {\it a} and {\it b} show the singular values. Usually the nonlinear
oscillations are presented by couple of nearly equal singular values. We can
recognize  three such oscillations in panels {\it a} and {\it b}
of Fig. 3. Another observation is that
the first four singular values are much larger than the other ones.
Thus we can conclude that the statistical dimension of the time series
is $S_{v}=S_{o}=4$. The panels {\it c} and {\it d}
present the percents of the total variance of the time
series contributed by the corresponding principal component.
As we can see almost all important information in the
investigated time series for the vestibular and for the optovestibular
nystagmus is concentrated in the first four principal components, i.e.
the change in the sensory input do not change the number of
important principal components. We can project the time series into the
subspace of the eigenvectors corresponding to the largest four to six
singular values. As a result we shall have time series which contain all
important information of the original time series and are free of the
influence of low-amplitude and high-frequency components such as noise.
The remaining panels of Fig. 3 present the first five principal components for the
time series for the vestibular nystagmus.
\par
As concluding remark we note that
the obtained in this article vestibular and optovestibular nystagmus time
series belong to the class of time series which are between the classes of
purely periodic and purely chaotic motions. The interests in investigation
of such time series
shows a steady increasing in the last years. The autocorrelations of the
time series show sign of both periodic and chaotic behavior and in addition
they present evidence for nonlinearity of the time series by the existence
of large local peaks of the amplitude after  region of exponential decreasing of the
peak amplitudes. The form of the histograms of the time series differs from
the Gaussian one and the most important frequencies in the power spectrum
are the same for both kinds of investigated nystagmus time series. The
singular spectrum analysis shows that the effects of low-amplitude and
high-frequency components of the time series can be easily eliminated by
projecting the series onto the subspace of four to six appropriate eigenvectors.
Finally, if stationary time series are result of purely linear process,
they can have
periodic, exponentially increasing or exponentially decreasing behavior in the
case of lack of noise. Other kind of behavior are evidence for presence of
noise or nonlinearity. Noise do not affect significantly the investigated time
series - several schemes for noise reduction did not changed significantly
the properties of the time series (as it is known the nonlinear characteristics
are very sensitive to presence of noise). The closeness of the values of the
various characteristic quantities when the eyes are open or closed shows that
the nystagmus time series are robust against changes in the sensory input
modality. 

\begin{flushleft}
{\large \bf \sc  references}
\end{flushleft}
{\scriptsize
\renewcommand{\baselinestretch}{1.0}
[$^{1}$]. WILSON V.J., G. M. JONES. 1979, Mammalian vestibular physiology.
Plenum Press, New York.
[$^{2}$]. KANTZ, H., T. SCHREIBER. 1997, Nonlinear Time Series Analysis.
Cambridge University Press, Cambridge.
[$^{3}$]. VITANOV, N.K., M. SIEFERT,
J. PEINKE.  Compt. rend. de l' Academie bulgare des Sciences, {\bf 55}, N 6,
2002, 15-20. 
[$^{4}$] VITANOV, N.K., M. SIEFERT, J. PEINKE, Ibid., {\bf 55},
N 9, 2002, 25-30.
[$^{5}$]. BALOH, R. W., J. Z. DEMER. Experimental Brain Research, {\bf 97},
1993, 334-342.
[$^{6}$]. FRASER A. M., H. L. SWINEY. Phys. Rev. A, {\bf 33}, 1986, 1134-
1140.
[$^{7}$]. HEGGER R., H. KANTZ. Phys. Rev. E, {\bf 60}, 1999, 4970-
4974.
[$^{8}$]. GRASSBERGER P., I. PROCACCIA. Physica D, {\bf 9}, 1983, 189-208.
[$^{9}$]. HEGGER R., H. KANTZ, T. SCHREIBER. CHAOS, {\bf 9}, 1999, 413-435.
[$^{10}$]. BROOMHEAD D. S., G. P. KING. Physica D, {\bf 20}, 1986, 217-236.
[$^{11}$]. VAUTARD R., P. YION, M. GHIL. Physica D, {\bf 58}, 1992, 95-126.
}\\
\vskip.5cm
Institute of Mechanics, Bulgarian Academy of Sciences, Akad. G. Bonchev Str.,
Bl. 4, 1113, Sofia, Bulgaria\\
e-mail: vitanov@imech.imbm.bas.bg
\newpage
\begin{flushleft}
{\large \bf \sc figure captions}
\end{flushleft}
\begin{enumerate}
\item {\bf Fig.1}\\ 
Time series and their images in the reconstructed phase space.
Panel {\it a}: vestibular nystagmus time series. Panel {\it b}: optovestibular
nystagmus time series. One unit for $t$ is $0.02$ s. One unit for $V$ is
$0.6$ degrees deviation of the eyes in the horizontal plane for panel
{\it a} and $0.9$ degrees deviation for panel {\it b}.
 Panel {\it c}: image of the time series in the subspace of the first 
three coordinates $X,Y,Z$
of the four-dimensional reconstructed space of the delay vectors for the case of
the vestibular nystagmus time series. Panel {\it d}: image of the time series
 in the subspace
of the first three coordinates $X.Y,Z$ of the four-dimensional reconstructed
space for the optovestibular time series. 
\item {\bf Fig.2}\\ 
Autocorrelations $A$, histograms $H$ and power spectra $P$
 for the
nystagmus time series. Panels {\it a,c,e}: quantities for the vestibular
nystagmus time series. Panels {\it b,d,e}: quantities for the optovestibular
nystagmus time series. One unit for $\Delta$ is $25$ Hz. 
\item {\bf Fig.3}\\ 
Results from the singular value decomposition and principal
component analysis of the time series. Panel {\it a}: singular values for
the vestibular nystagmus time series. Panel {\it b}: singular values for
the optovestibular nystagmus time series. Panel {\it c}: Portion of the
total variance, due to the corresponding 
principal components for the vestibular nystagmus. Panel {\it d}: Portion
of the total variance of the time series due to the corresponding 
principal components for the optovestibular nystagmus. Panel {\it e}:
First principal component for the vestibular nystagmus. Panel {\it f}: 
second principal component for the vestibular nystagmus time series.
Panel {\it g}: third principal component for the vestibular nystagmus time 
series. Panel {\it h}: fourth principal component for the vestibular
nystagmus time series. Panel {\it i}: fifth principal component for the
vestibular nystagmus time series. 
\end{enumerate}
\newpage
\begin{figure}[t]
\vskip-4cm
\includegraphics[angle=-90,totalheight=12cm]{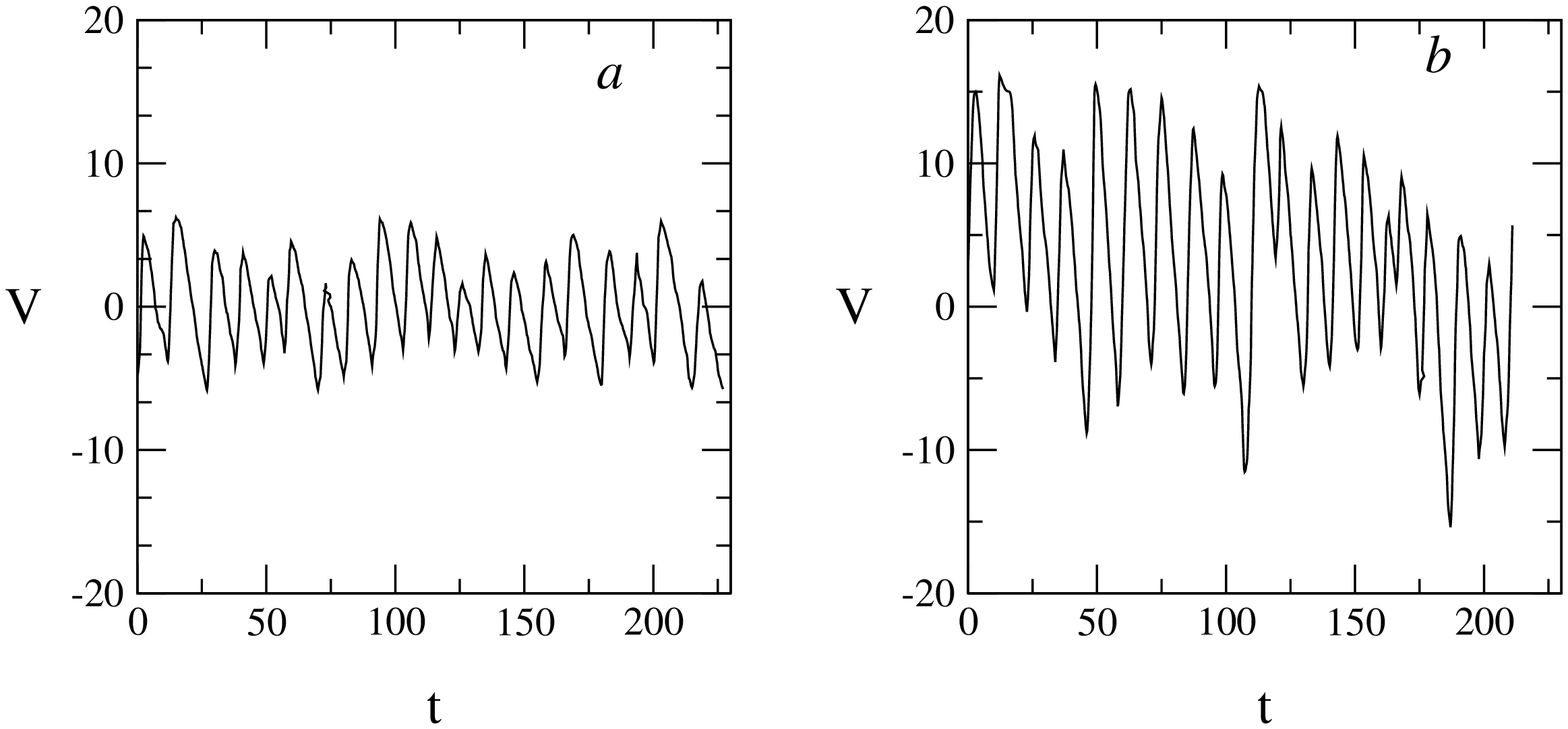}
\vskip-4cm
\includegraphics[angle=-90,totalheight=5.5cm]{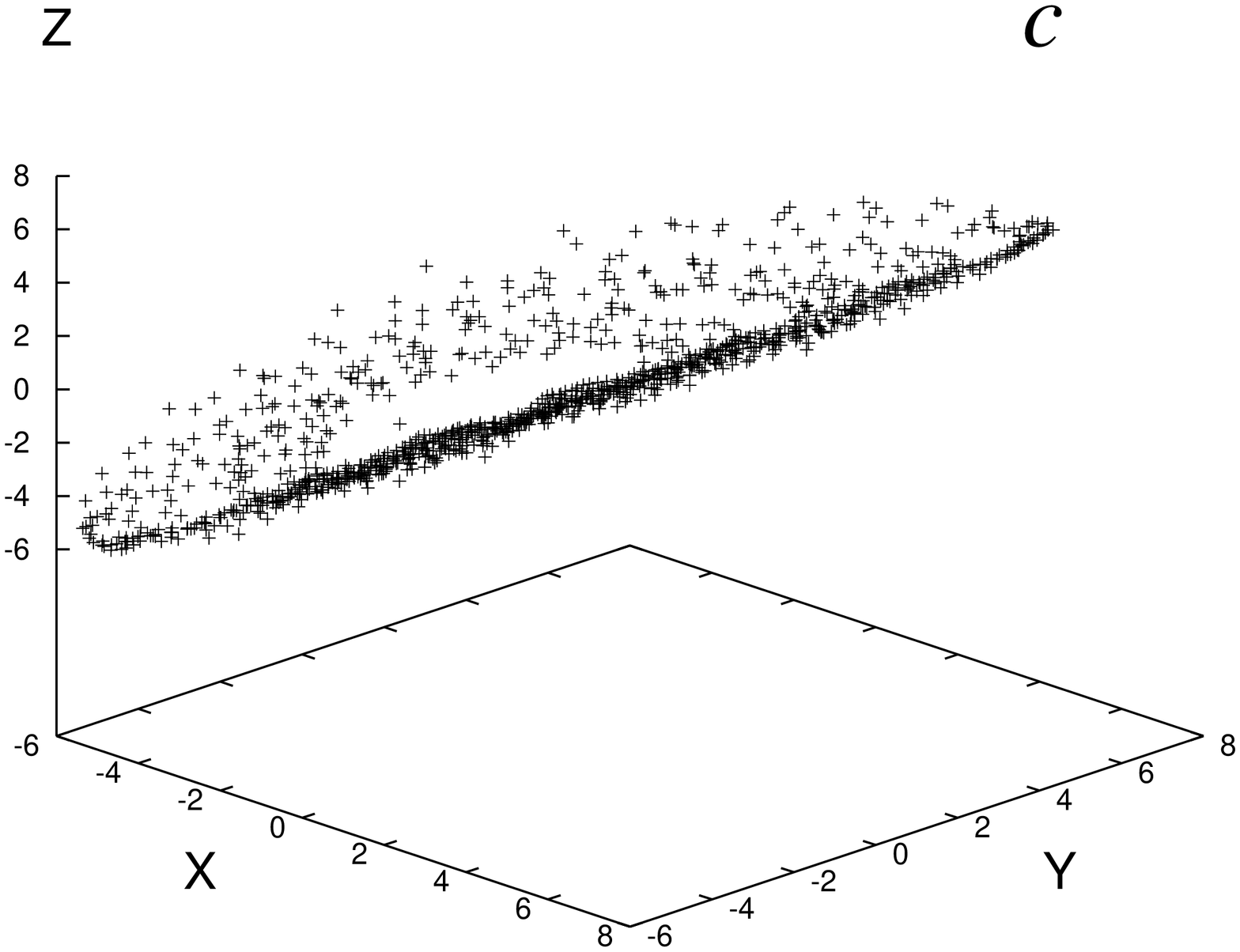}
\includegraphics[angle=-90,totalheight=5.5cm]{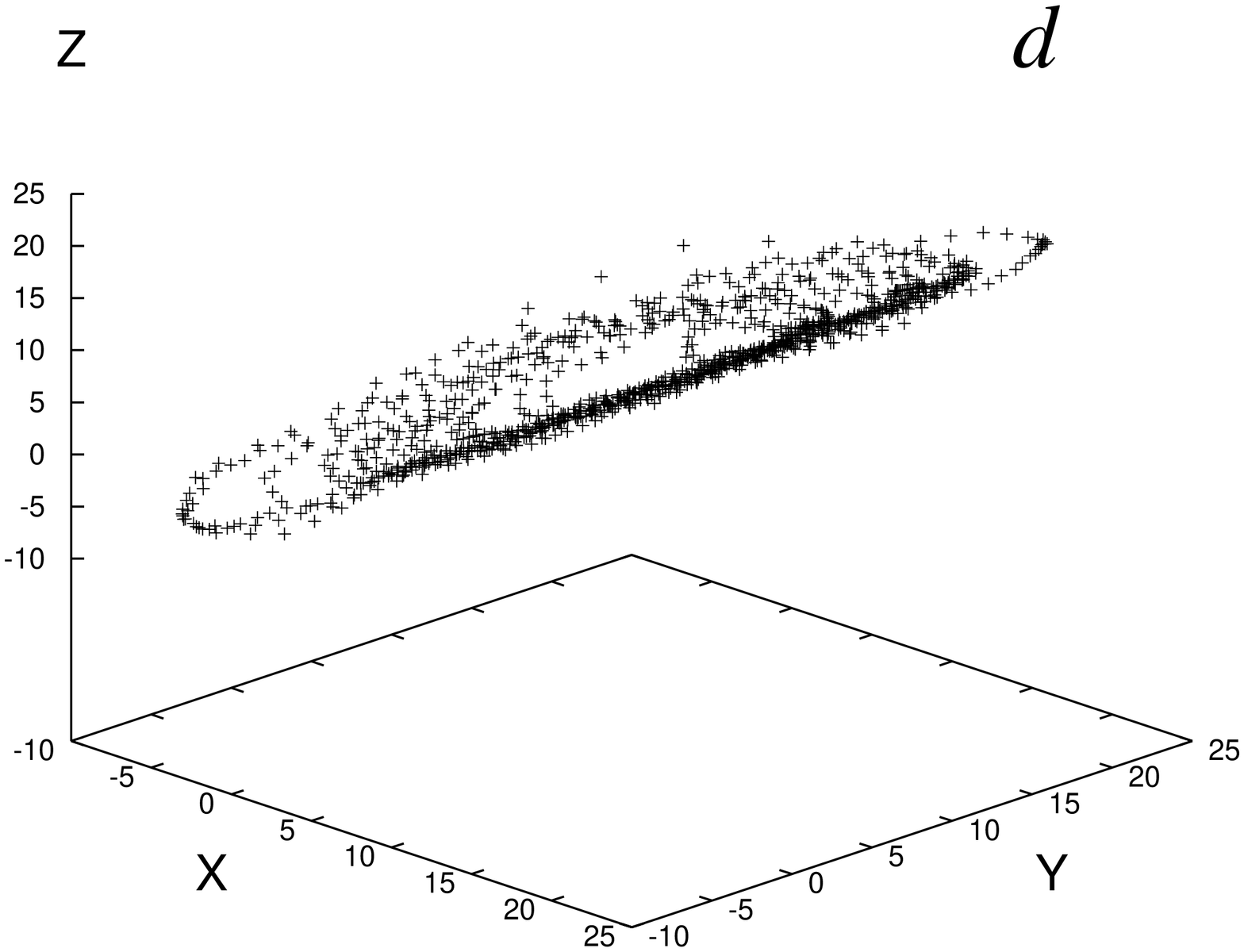}
\end{figure}
\begin{figure}[t]
\includegraphics[angle=-90,totalheight=12cm]{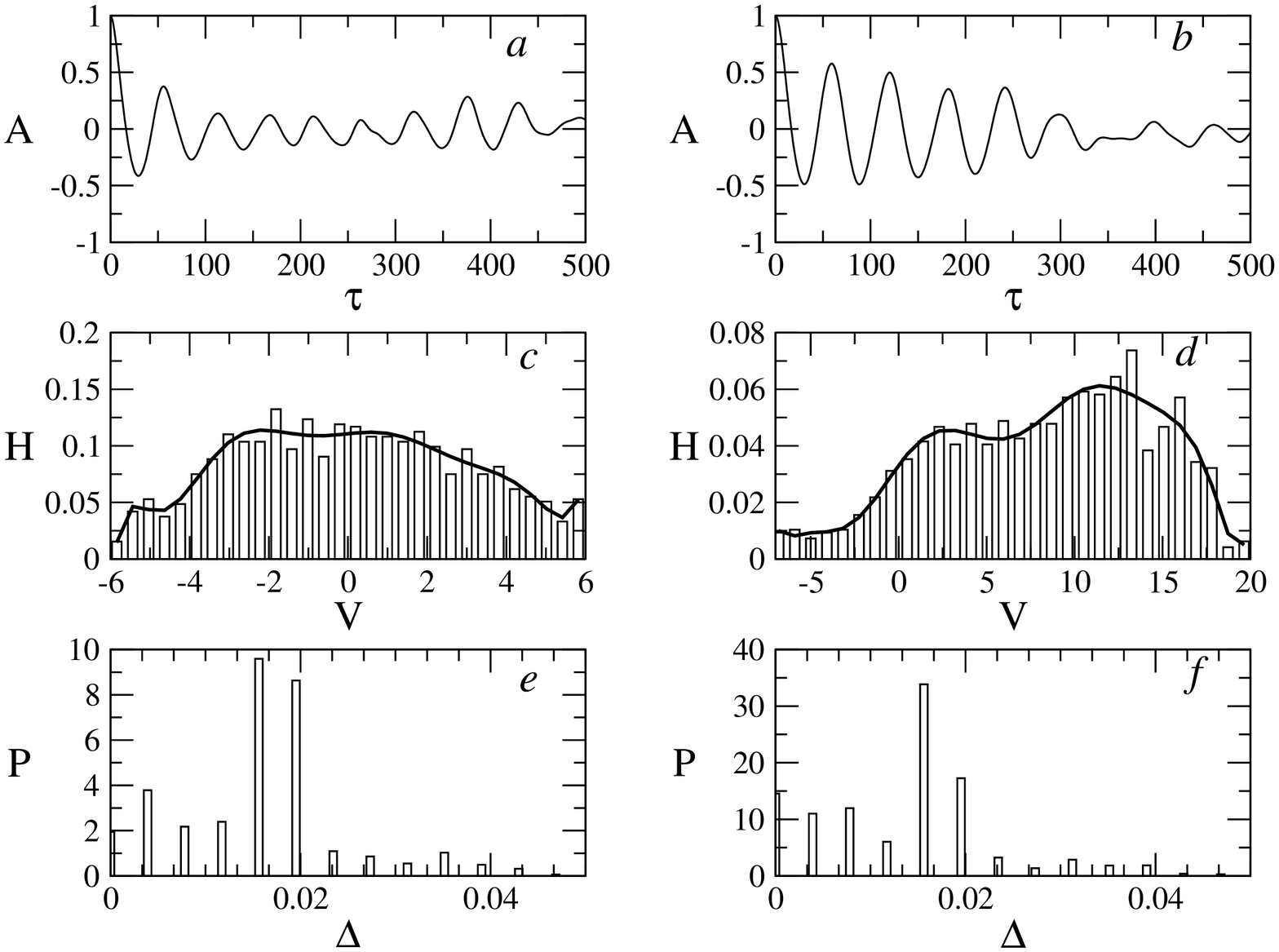}
\end{figure}
\begin{figure}[t]
\includegraphics[angle=-90,totalheight=12cm]{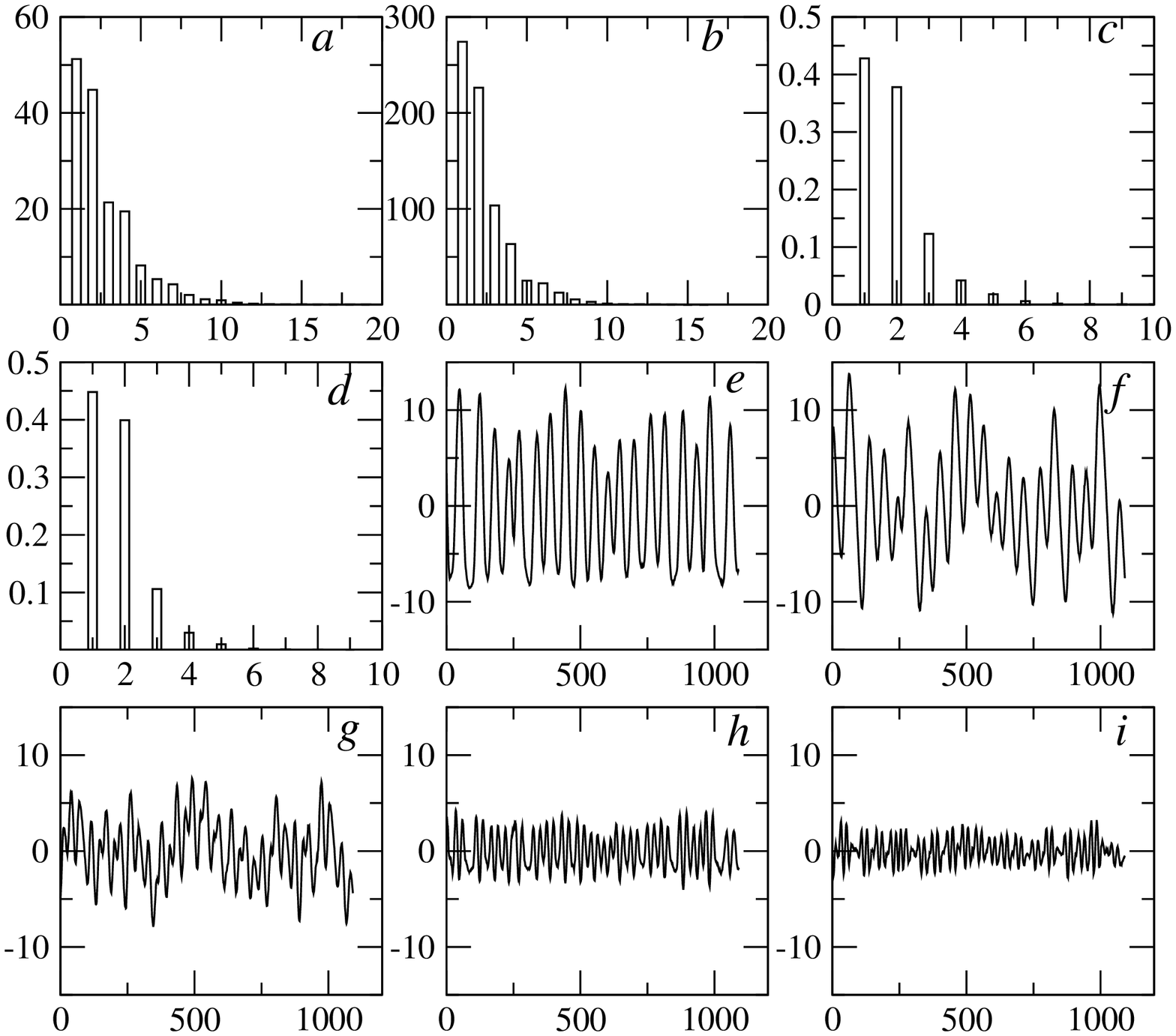}
\end{figure}
\end{document}